\documentclass[12pt,reqno]{amsart}
\usepackage{amsmath,amsfonts,amsthm,amssymb,dsfont,mathtools}
\newcommand\version{November 23, 2026}
\usepackage{amsxtra}

\newtheorem{theorem}{Theorem}

\newtheorem{lemma}{Lemma}

\theoremstyle{definition}

\theoremstyle{remark}
\newtheorem{remark}[theorem]{Remark}

\numberwithin{equation}{section}

\newcommand{\C}{\mathbb{C}}

\newcommand{\eps}{\epsilon}
\renewcommand{\epsilon}{\varepsilon}

\renewcommand{\phi}{\varphi}
\newcommand{\R}{\mathbb{R}}

\newcommand\id{\mathds{1}}

\DeclareMathOperator{\dom}{dom}

\DeclareMathOperator{\infspec}{inf\, spec\,}

\renewcommand\hat\widehat

\begin{document}

\title[Point-interacting bosons in the plane]{Large particle number asymptotics for point-interacting bosons in the plane}

\author{Rupert L. Frank}
\address{Rupert L. Frank, Mathematisches Institut, Ludwig-Maximilians Universit\"at M\"un\-chen, Theresienstr. 39, 80333 M\"unchen, Germany, and Munich Center for Quantum Science and Technology, Schellingstr. 4, 80799 M\"unchen, Germany}
\email{r.frank@lmu.de}

\author{Robert Seiringer}
\address{Robert Seiringer, Institute of Science and Technology Austria, Am Campus 1, 3400 Klosterneuburg, Austria}
\email{rseiring@ist.ac.at}

\thanks{\copyright\, 2026 by the authors. This paper may be  
reproduced, in
its entirety, for non-commercial purposes.}

\begin{abstract} We consider $N$ bosons in two dimensions interacting via point interactions, modeled by the TMS Hamiltonian. We prove that their ground state energy has the asymptotic behavior $-e^{C N }$ for large $N$, for an explicit constant $C$ related to the sharp constant in the Ladyzhenskaya--Gagliardo--Nirenberg inequality. Such asymptotic behavior was predicted in the physics literature, based on a variant of the mean-field approximation.  
\end{abstract}

\date{\version}

\maketitle


\section{Introduction and Main Results}

\subsection{Introduction} Point interactions  play a prominent role in quantum physics, due to their universal nature. They have been predicted to be relevant effective models for cold atomic gases \cite{HS,AV}. In this paper, we shall consider point-interacting bosons in the two-dimensional plane. These are described by the TMS Hamiltonian \cite{tms,DA}, which will be recalled in the next section. It is well known that this Hamiltonian is bounded from below for any particle number $N$, but the best known lower bounds on its ground state energy behave as $- e^{C N^2}$ for a suitable constant $C>0$ \cite{DA,DR}. On the other hand, a mean-field approximation predicts the asymptotic behavior $- e^{CN}$ \cite{R,HS} for an explicit $C$ involving the sharp constant $a_*$ in the Ladyzhenskaya--Gagliardo--Nirenberg inequality 
\begin{equation}\label{LGN}
\| \nabla u\|_2^2 \|u\|_2^2 \geq \frac{a_*}2 \, \|u\|_4^4  \quad , \quad u\in H^1(\R^2) \,.
\end{equation}
For more on this inequality, including the numerical value $a_* \approx 11.70$, see \cite{W}.

In this paper, we shall prove that the prediction from \cite{R,HS} is indeed correct.

\subsection{Asymptotics of ground state energy} Let $\mathfrak{e}_N$ denote the ground state energy of the TMS Hamiltonian defined in \eqref{Halpha} below. The latter depends on a parameter $\alpha\in \R$ determining the two-body binding energy $\mathfrak{e}_2$, but the ratio $\mathfrak{e}_N/\mathfrak{e}_2$ is independent of $\alpha$, and only depends on $N$. It has the following asymptotic behavior.

\begin{theorem} \label{thm:main}
Let $\mathfrak{e}_N$ denote the ground state energy of the TMS Hamiltonian defined in \eqref{Halpha}, and $a_*$ the sharp constant in \eqref{LGN}. Then 
\begin{equation}
\lim_{N\to \infty} \frac 1N \ln \frac{\mathfrak{e}_N}{\mathfrak{e}_2}  =  \frac{ 8 \pi}{a_*} \,.
\end{equation}
\end{theorem}

The proof of this theorem will be given in Sections~\ref{ub} (upper bound) and~\ref{lb} (lower bound). 

These bounds will be carried out in the charge formulation of the ground state energy problem, defined on functions $\xi$ on $\R^{2(N-1)}$ that are permutation symmetric in the last $N-2$ variables. For the upper bound, following \cite{R}, we take $\xi$ to be a product state in these last $N-2$ variables, multiplied by an appropriately chosen function of the remaining variable. This choice of product states is reminiscent of the Hartree approximation in the mean field limit, which was studied, in a rather general setting, in \cite{LNR}. In the charge formulation, however, the effective Hamiltonian does \emph{not} consist only of one-body and two-body terms, and consequently the approach from \cite{LNR} is not applicable. For the lower bound we instead proceed by an ad-hoc method, where at a crucial point a weak form of a result of \cite{JV} enters. Interestingly, the proof in \cite{JV} relies on the De Finetti theorem for product observables with logarithmic dependence on the dimension of the one-particle space \cite{logdF1,logdF2} (see also \cite[App.~A]{R1}), which is a variant of the De Finetti theorem lying at the basis of \cite{LNR}. For earlier results on mean-field limits using this approach, we refer to \cite{R1,R2}.

\subsection{Universality for weakly interacting bosons}

The point interaction model arises naturally for systems with genuine short range interactions in the weak coupling limit. 
Consider $N$ bosons in $\R^2$, with Hamiltonian
\begin{equation}
H = \sum_{j=1}^N \left( - \Delta_j \right) - \lambda \sum_{1\leq i<j\leq N} V(x_i-x_j)
\end{equation}
in $L^2_{\rm sym}(\R^{2N})$, where $V\in L^1(\R^2)$ with $\int V > 0$ is even and sufficiently regular (e.g. $V \in L^2(\R^2)$ and $ \int_{\R^2} |V(x)| (1+|x|)^p dx < \infty$ for $p>0$ suffice).
Let $E_0(N,\lambda)$ denote the ground state energy of $H$ for $\lambda > 0$. The following theorem shows  universality in the weak coupling limit, and the appearance of point interactions as an effective model. 

\begin{theorem} \label{thm:2}
With the definitions above, 
\begin{equation}
\lim_{\lambda\to 0} \frac{E_0(N,\lambda)} {E_0(2,\lambda)} = \frac{\mathfrak{e}_N}{\mathfrak{e}_2} \,,
\end{equation}
independently of $V$. 
\end{theorem}

As we shall explain momentarily, this theorem can be deduced in a straightforward way from the analysis in \cite{GH}. We remark that it follows from \cite{S,BCEZ} 
that
\begin{align}\nonumber
&E_0(2,\lambda ) \\ & = - \exp\left( - \frac {8\pi}{\lambda \int V}  - \frac 2 {(\int V)^2} \int_{\R^4} V(x) \ln|x-y| V(y) \, dx dy  - 2 \gamma_{\rm E} + \ln 8 +   o(1)_{\lambda\to 0} \right),
\end{align}
where $\gamma_{\rm E}$ denotes Euler's constant.

Let us give the details of how to deduce Theorem \ref{thm:2} from the results of \cite{GH}. Their assertion is that
\begin{equation}
	\label{eq:hamrescaled}
	\sum_{j=1}^N(-\Delta_j) - \epsilon^{-2} \frac{4\pi}{\int V} \frac{1}{\ln(1/\epsilon)} \, \sum_{1\leq i<j\leq N} V((x_i-x_j)/\epsilon)
\end{equation}
converges, as $\epsilon\to 0_+$, in norm resolvent sense to the TMS Hamiltonian with an explicit parameter $\alpha\in\R$; see \cite[Eq.~(4.34)]{GH}. While the value of $\alpha$ is irrelevant for us, what is crucial is that it is independent of $N$. Thus, the bottom of the spectrum of the operator \eqref{eq:hamrescaled} is given by $\mathfrak{e}_N + o(1)$ as $\epsilon\to 0$. By scaling, it follows that the bottom of the spectrum of the operator
\begin{equation}
\sum_{j=1}^N(-\Delta_j) - \frac{4\pi}{\int V} \frac{1}{\ln(1/\epsilon)} \, \sum_{1\leq i<j\leq N} V(x_i-x_j)
\end{equation}
is given by $\epsilon^2 \mathfrak{e}_N + o(\epsilon^2)$. Applying this fact with $
\epsilon = \exp\left( - 4\pi/ (\lambda \int V) \right)$ proves Theorem \ref{thm:2}.


\section{TMS Hamiltonian}

The purpose of this section is two-fold. First, we introduce the TMS Hamiltonian, which describes $N\geq 2$ bosons of mass $1/2$ in two dimensions, interacting via zero range forces, as a self-adjoint operator in the Hilbert space $L^2_{\rm sym}(\R^{2N})$ of permutation symmetric functions. Its ground state energy $\mathfrak{e}_N$ (that is, the bottom of the spectrum of this operator) is the main quantity of interest. Second, we introduce an `effective' operator, acting in the Hilbert space
\begin{equation}
\mathcal H:= L^2(\R^2) \otimes L^2_{\rm sym}(\R^{2(N-2)})
\end{equation}
of functions on $\R^{2(N-1)}$ that are symmetric in the last $N-2$ variables. The elements of $\mathcal H$ are often called charges in the physics literature. The bottom of the spectrum of the effective operator in $\mathcal H$, denoted by $t_N$, is directly related to the quotient $\mathfrak{e}_N/\mathfrak{e}_2$. In the following sections we will study the asymptotics of $t_N$.

Let us recall the definition of the TMS Hamiltonian, following \cite{tms,DA}. Let $\mu>0$. In the Hilbert space $\mathcal H$ we introduce the quadratic form
\begin{align}\nonumber 
	T_\mu (\xi) & =   \frac \pi 2 \int | \widehat \xi (q, \hat p_{12}) |^2 \ln \left( \tfrac 12 q^2 + | \hat p_{12}|^2 + \mu \right) dq dp_3 \cdots dp_N 
	\\ \nonumber & \quad  -  {2(N-2)} \int \frac{ \widehat \xi(p_1+p_2, \hat p_{12})^* \widehat \xi(p_1+ p_3, \hat p_{13} )}{ |\vec p|^2 + \mu} \, dp_1 \cdots dp_N
	\\ & \quad  -  \frac{(N-2)(N-3)}2 \int \frac{ \widehat \xi(p_1+p_2, \hat p_{12})^* \widehat \xi(p_3+ p_4, \hat p_{34} )}{ |\vec p|^2 + \mu} \, dp_1 \cdots dp_N \,.
	\label{def:T}
\end{align}
Here, for $\vec p = (p_1,\dots,p_N) \in \R^{2N}$ we denote $\hat p_{ij} = (p_1,\dots, \not \!\! p_i, \dots , \not\!\! p_j, \dots, p_N)\in\R^{2(N-2)}$, and $z^*$ denotes the complex conjugate of a number $z\in\C$. The domain of this quadratic form, $\dom T_\mu$, is given by those $\xi\in\mathcal H$ for which the first term on the right side is finite. The second and third terms are bounded quadratic forms in $\mathcal H$, as we will see later on.

Now let $\alpha\in\R$ and define a quadratic form $\mathcal E^\alpha$ in $L^2_{\rm sym}(\R^{2N})$ as follows. We fix a parameter $\mu>0$ (later, we will see that the construction is independent of this parameter) and define the domain of $\mathcal E^\alpha$ to be
\begin{equation}
	\dom \mathcal E^\alpha = \left\{ \psi \in L^2_{\rm sym}(\R^{2N}) :\
	\psi = \phi + G_\mu \xi \ \text{with}\ \phi \in H_{\rm sym}^{1}(\R^{2N}) \ \text{and}\ \xi\in\dom T_\mu \right\}\,.
\end{equation}
Here $G_\mu \xi$ is defined through its Fourier transform
\begin{equation}
\widehat{G_\mu \xi}(p_1,\dots,p_N) = \frac 1{ |\vec p|^2 + \mu} \sum_{i<j} \widehat\xi(p_i + p_j, \hat p_{ij} ) \,.
\end{equation}
Since $G_\mu\xi\not\in H^1_{\rm sym}(\R^{2N})$ for all $\xi\neq 0$, the decomposition $\psi = \phi + G_\mu \xi$ of $\psi\in\dom\mathcal E^\alpha$ is unique and we can set
\begin{equation}\label{Halpha}
\mathcal{E}^\alpha(\psi) = - \mu \|\psi\|^2 + \| \nabla \varphi\|^2 + \mu \|\varphi\|^2 + \frac {N(N-1)}2  \left(  T_\mu ( \xi) + \alpha \|\xi\|^2 \right).
\end{equation}
One easily checks that the definition of $\dom\mathcal E^\alpha$ and the value of $\mathcal E^\alpha(\psi)$ are independent of $\mu$. (To do so, one verifies that $\xi$ is uniquely determined by $\psi$, independently of $\mu$.)

As shown in \cite{DA}, the quadratic form $\mathcal E^\alpha$ is bounded from below and closed in the Hilbert space $L^2_{\rm sym}(\R^{2N})$ and, therefore, generates a self-adjoint operator. The bottom of its spectrum is
\begin{equation}
\mathfrak{e}_N := \inf_{0\neq \psi\in\dom\mathcal E^\alpha} \frac{\mathcal E^\alpha(\psi)}{\|\psi\|^2} \,.
\end{equation}
One easily sees that $\mathfrak{e}_N$ is the largest $-\mu$ such that $T_\mu(\xi) + \alpha \|\xi\|^2$ is non-negative. In particular, for $N=2$, we see that 
$\mathfrak{e}_2 = - \exp(-2\alpha/\pi)$. By scaling, $\mathfrak{e}_N/ \mathfrak{e}_2$ is independent of $\alpha$; 
in fact, we have the identity
\begin{equation}
\mathfrak{e}_N = \mathfrak{e}_2 \, e^{-\frac 2 \pi t_N} \,,
\end{equation}
where
\begin{equation}\label{def:tN}
 t_N =  \inf_{0\neq\xi\in\dom T_\mu} \frac{T_\mu(\xi)}{\|\xi\|^2} - \frac \pi 2 \ln \mu =  \inf_{0\neq\xi\in\dom T_1} \frac{T_1(\xi)}{\|\xi\|^2} \,.
\end{equation}
The second equality follows by scaling.

In view of these facts, Theorem~\ref{thm:main} is equivalent to the statement
\begin{equation}
\lim_{N\to \infty} \frac 1N t_N = - \frac{(2\pi)^2}{a_*} \,,
\end{equation}
which we shall prove in the following. It will be convenient to introduce the operators $S_0$, $S_1$ and $S_2$ in $\mathcal H$, corresponding to the three terms in \eqref{def:T} at $\mu=1$, in such a way that 
\begin{equation}\label{def:S}
T_1(\xi) = \langle \xi | S_0 | \xi \rangle -2 (N-2) \langle \xi | S_1 | \xi \rangle - \frac{(N-2)(N-3)}2\langle \xi | S_2| \xi \rangle \, .
\end{equation}  
For computing its infimum, we shall use that it clearly suffices to consider functions $\xi$ with non-negative Fourier transform, $\hat \xi \geq 0$. 


\section{Upper bound} \label{ub}

Our upper bound on $t_N$ in \eqref{def:tN} is based on a product ansatz, following \cite{R}. We shall pick a normalized $\xi\in\mathcal H$ such that 
 $\widehat \xi \geq 0$. Then the second term in \eqref{def:S} is non-positive, 
\begin{equation}
- \langle \xi | S_1 | \xi\rangle =  -   \int \frac{ \widehat \xi(p_1+p_2, \hat p_{12}) \widehat \xi(p_1+ p_3, \hat p_{13} )}{ |\vec p|^2 + 1} dp_1 \cdots dp_N \leq 0 \,.
\end{equation}
Moreover, by Jensen's inequality
\begin{align}\nonumber
- \langle \xi | S_2 | \xi\rangle &= -  \int \frac{ \widehat \xi(p_1+p_2, \hat p_{12}) \widehat \xi(p_3+ p_4, \hat p_{34} )}{ |\vec p|^2 + 1} dp_1 \cdots dp_N
\\ & \leq - \frac { \left(  \int  \widehat \xi(p_1+p_2, \hat p_{12}) \widehat \xi(p_3+ p_4, \hat p_{34} )  dp_1 \cdots dp_N \right)^2 } { \int  \widehat \xi(p_1+p_2, \hat p_{12}) \widehat \xi(p_3+ p_4, \hat p_{34} ) \left( |\vec p|^2 + 1 \right) dp_1 \cdots dp_N} \,.
\end{align}
Likewise,
\begin{align}\nonumber
\frac 2\pi  \langle \xi | S_0 | \xi\rangle &=  \int | \widehat \xi (q, \hat p_{12}) |^2 \ln \left( \tfrac 12 q^2 + | \hat p_{12}|^2 + 1 \right) dq dp_3 \cdots dp_N 
\\  & \leq   \ln \left( \int | \widehat \xi (q, \hat p_{12}) |^2 \left( \tfrac 12 q^2 + | \hat p_{12}|^2 + 1 \right) dq dp_3 \cdots dp_N \right).
\end{align}

 We shall make the ansatz
\begin{equation}\label{ts}
\widehat\xi(q,p_1,\dots,p_{N-2}) = \widehat v(q) \prod_{j=1}^{N-2} \widehat u(p_j)
\end{equation}
for $L^2$-normalized $u,v\in H^1(\R^2)$ such that $\widehat u, \widehat v\geq 0$. Then 
\begin{equation}
 \int | \widehat \xi (q, \hat p_{12}) |^2 \left( \tfrac 12 q^2 + | \hat p_{12}|^2 + 1 \right) dq dp_3 \cdots dp_N = 1 + \tfrac 12 \| \nabla v\|^2 + (N-2)  \| \nabla u\|^2 \,.
\end{equation}
Moreover,
\begin{equation}
 \int  \widehat \xi(p_1+p_2, \hat p_{12}) \widehat \xi(p_3+ p_4, \hat p_{34} )  dp_1 \cdots dp_N = \left( \int   \widehat v(p_1+p_2) \widehat u(p_1)\widehat u(p_2) dp_1 dp_2 \right) ^2
\end{equation}
and
\begin{align}\nonumber 
& \int  \widehat \xi(p_1+p_2, \hat p_{12}) \widehat \xi(p_3+ p_4, \hat p_{34} )  |\vec p|^2 \, dp_1 \cdots dp_N 
\\ & = (N-4) \| \nabla u\|^2  \left( \int \widehat  v(p_1+p_2)\widehat u(p_1)\widehat u(p_2) dp_1 dp_2 \right) ^2 \nonumber
\\ & \quad + 4 \left( \int \widehat  v(p_1+p_2)\widehat u(p_1)\widehat u(p_2) dp_1 dp_2 \right) \left( \int \widehat  v(p_1+p_2)\widehat u(p_1)\widehat u(p_2) |p_1|^2 \, dp_1 dp_2 \right) 
\end{align}
We thus conclude that 
\begin{equation}
\limsup_{N\to \infty} \frac 1N T_1(\xi) \leq - \frac 12 \frac{ \left(  \int\widehat v(p_1+p_2)\widehat u(p_1)\widehat u(p_2) dp_1 dp_2 \right)^2 }{ \| \nabla u \|^2  } \,.
\end{equation}
Since
\begin{equation}
 \int\widehat v(p_1+p_2)\widehat u(p_1)\widehat u(p_2) dp_1 dp_2 = 2\pi \int v^* u^2 \,,
\end{equation}
the optimal choice of $v$, given $u$, is  $v = u^2 /\|u\|_4^2$, and, in view of \eqref{def:tN}, this gives the desired upper bound
\begin{equation}
\limsup_{N\to \infty} \frac 1 N t_N \leq  - \frac{(2\pi )^2}{a_*} \,.
\end{equation}
Here we used the fact that the sharp constant \(a_*\) in \eqref{LGN} remains the same when restricted to functions with nonnegative Fourier transform: indeed, replacing any function \(u\) by \(\tilde u=\mathcal F^{-1}|\widehat u|\) preserves its \(L^2\) norm and Dirichlet integral, while \(|\tilde u|_4\ge|u|_4\) by Plancherel’s theorem and the pointwise inequality \(|\widehat u*\widehat u|\le|\widehat u|*|\widehat u|\). Moreover, for an optimizer $u$, $u^2 \in H^1(\R^2)$.


\section{Lower bound} \label{lb}

The proof of the lower bound is substantially more involved. We shall split it into several steps.

\subsection{Step 1. Reduction to $S_2$} 
We shall first show that it suffices to consider the third term $S_2$ in \eqref{def:S}. 
For the first term, we simply have
$S_0 \geq 0$. 
For the second, we use the following lemma.

\begin{lemma} \label{lem:1}
For any $0< \beta < 1$, $\lambda \in \R$  and $\psi \in L^2(\R^2)$, 
\begin{equation}\label{eq:lem1}
0\leq \int_{\R^4} \frac{  \psi(x)^* \psi(y)}{|x|^2 + |y|^2 + \lambda^2} \, dx dy \leq  \frac {\pi^2}{ \sin \pi \beta} \int_{\R^2} |\psi(x)|^2 \left( \frac { |x|^2}{|x|^2 + \lambda^2}  \right)^{\beta}dx \,.
\end{equation}
\end{lemma}

\begin{proof}
	By writing $(|x|^2+|y|^2+\lambda^2)^{-1} = \int_0^\infty e^{-t(|x|^2+|y|^2+\lambda^2)}\,dt$, we obtain the left inequality in \eqref{eq:lem1}. By Cauchy--Schwarz, we have for any positive function $h$,
\begin{equation}
\int_{\R^4} \frac{ \psi(x)^* \psi(y)}{|x|^2 + |y|^2 + \lambda^2} \, dx dy \leq  \int_{\R^2} |\psi(x)|^2 \frac 1{h(x)} \int_{\R^2} \frac{h(y)}{|x|^2 + |y|^2 + \lambda^2} \,dy dx \,.
\end{equation}
We choose $h(y) = |y|^{-2\beta}$. This implies the claim by scaling and the fact that
\begin{equation}
\int_{\R^2} \frac{|z|^{-2\beta}}{1+|z|^2}\,dz = 2\pi \int_0^\infty \frac{r^{-2\beta+1}\,dr}{1+r^2} = \pi \int_0^\infty \frac{s^{-\beta}\,ds}{1+s} = \frac{\pi^2}{\sin\pi\beta} \,.
\end{equation}
For the last integral we have substituted $t = \frac{s}{1+s}$ to arrive at a beta function, which we have simplified using Euler's reflection formula for the gamma function.
\end{proof}

Shifting the integration variable $p_1$ to $p_1- p_2-p_3$, we can write
\begin{equation}
\langle \xi | S_1| \xi\rangle  = \int \frac{ \widehat \xi(p_1-p_3, \hat p_{12})^* \widehat \xi(p_1- p_2, \hat p_{13} )}{ | p_1 - p_2-p_3|^2  + \sum\nolimits_{j=2}^N | p_j|^2 + 1}  \, dp_1 \cdots dp_N
\end{equation}
and hence
\begin{equation}
|\langle \xi | S_1 |\xi\rangle|   \leq  \int \frac{ | \widehat \xi(p_1-p_3, \hat p_{12})|  |\widehat \xi(p_1- p_2, \hat p_{13} )| }{  |p_2|^2 + |p_3|^2 + \sum\nolimits_{j=4}^N | p_j|^2 } \, dp_1 \cdots dp_N \,.
\end{equation}
Applying Lemma~\ref{lem:1} for the integration over the variables $p_2$ and $p_3$, we obtain, for normalized $\xi\in\mathcal H$, 
\begin{align}\nonumber
|\langle \xi | S_1 |\xi\rangle|   &\leq  \frac {\pi^2}{ \sin \beta \pi}\int  | \widehat \xi(p_1-p_3, \hat p_{12})|^2 \left( \frac{ |p_3|^2  }{  |p_3|^2 + \sum\nolimits_{j=4}^N | p_j|^2 }\right)^\beta dp_1 dp_3 \cdots dp_N
\\ \nonumber & = \frac {\pi^2}{ \sin \beta \pi}\int  | \widehat \xi(p_1, \hat p_{12})|^2 \left( \frac{ |p_3|^2  }{   \sum\nolimits_{j=3}^N | p_j|^2 }\right)^\beta dp_1 dp_3\cdots dp_N
\\ \nonumber & \leq \frac {\pi^2}{ \sin \beta \pi} \left( \int  | \widehat \xi(p_1, \hat p_{12})|^2  \frac{ |p_3|^2  }{   \sum\nolimits_{j=3}^N | p_j|^2 } dp_1 dp_3\cdots dp_N \right)^\beta \\
& = \frac {\pi^2}{ \sin \beta \pi} (N-2)^{-\beta}.
\end{align}
In the next-to-last step, we used Jensen's inequality and, in the last step, the bosonic symmetry of $\xi$ in the last $N-2$ variables. Optimizing over the choice of $\beta$ thus yields $\|S_1\| \leq O(N^{-1} \ln N)$. 

\begin{remark} Using a product trial state similarly as in the last section, one can show that the $N$-dependence in the obtained bound $\|S_1\| \leq O(N^{-1} \ln N)$ is actually optimal. 
\end{remark}

\subsection{Step 2. Energy localization}  
The operator $K = e^{\frac 2\pi S_0}$ in $\mathcal H$ is given by the Fourier multiplier $K(q,p_3,\dots,p_N)=\tfrac 12 q^2 + | \hat p_{12}|^2 + 1$. Let us write 
\begin{equation}\label{eq:s2decomp}
 \left\langle \xi \left| S_2   \right| \xi \right\rangle = 
\int \frac{ \widehat \xi(p_1+p_2, \hat p_{12})^* \widehat \xi(p_3+ p_4, \hat p_{34} )}{ |\vec p|^2 + 1}  \left(  \id_R + \id_R^\perp  \right)  dp_1 \cdots dp_N
\end{equation}
with $\id_R$ the characteristic function of $|\ln(K(p_1+p_2,\hat p_{12})/K(p_3+p_4,\hat p_{34}))| \leq R$ for some $R>0$. The term involving $\id_R^\perp=\id-\id_R$ can be bounded with the aid of the following lemma.

\begin{lemma} \label{lem:2}
For any $R>0$, $\lambda \geq 0$  and $\psi \in L^2(\R^2)$, 
\begin{equation}
\left| \int_{\R^4} \frac{  \psi(x)^* \psi(y)}{|x|^2 + |y|^2 + \lambda^2} \id \big( | \ln (|x|^2+\lambda^2)/(|y|^2+\lambda^2) | > R\big) \,dx dy \right| \leq  C e^{-R/2} \| \psi\|^2 \,. 
\end{equation}
\end{lemma}

\begin{proof}
Without loss of generality, we can restrict our attention to non-negative radial functions $\psi$. Introducing logarithmic coordinates $|x|^2+\lambda^2 = e^u$, $|y|^2+\lambda^2 = e^v$, and bounding the denominator by $|x|^2+|y|^2+\lambda^2 \geq \frac 12 (|x|^2+|y|^2 + 2\lambda^2)$, the desired bound follows from
\begin{equation}
\left| \int_{\R^2} \frac{  f(u)^* f(v)}{\cosh\big( \frac 12(u-v)\big)} \id \big( | u-v| > R\big) \, du dv \right| \leq  C e^{-R/2} \| f\|^2 
\end{equation}
for functions $f\in L^2(\R)$, which is a simple consequence of the Cauchy--Schwarz inequality. 
\end{proof}

We shall apply this lemma to the integration over $p_1-p_2$ and $p_3-p_4$, with the result that 
\begin{equation} 
\int \frac{ \widehat \xi(p_1+p_2, \hat p_{12})^* \widehat \xi(p_3+ p_4, \hat p_{34} )}{ |\vec p|^2 + 1}   \id_R^\perp  \, dp_1 \cdots dp_N \leq C e^{-R/2} \| \xi\|^2 \,.
\end{equation}

To bound the term involving $\id_R$ in \eqref{eq:s2decomp}, we pick a nonnegative $\chi\in C_c^\infty((-1,1))$ with
$\int\chi^2=1$ (independent of $R$). For $L>0$ and $s \in \R$,   define the Fourier multipliers  
\begin{equation}
	\chi_s=L^{-1/2}\chi\left( (\ln K-s)/L\right) 
\end{equation}
satisfying $\int_\R\chi_s^2 \, d s= \id$. For $\vec p$ in the support of $\id_R$, we shall use that
\begin{align}\nonumber
& 1 - \int_\R \chi_s(p_1+p_2,\hat p_{12} ) \chi_s(p_3+p_4,\hat p_{34}) \, ds  \\ \nonumber 
& =\frac12\left\| \chi-\chi\big(\,\cdot+ L^{-1} \ln (K(p_1+p_2,\hat p_{12})/K(p_3+p_4,\hat p_{34})) \big)\right\|_2^2 
\\
& \leq C L^{-2}|\ln(K(p_1+p_2,\hat p_{12})/K(p_3+p_4,\hat p_{34}))|^2 \leq C (R/L)^2 \,. 
\end{align}

For $\xi$ with $\hat \xi\geq 0$ we thus conclude that, if $R/L$ is sufficiently small,
\begin{equation}\label{35}
 \left\langle \xi \left| S_2  \right| \xi \right\rangle \leq \left( 1 - C(R/L)^2\right)^{-1}  \left\langle \xi \left| \int_\R\chi_s S_2 \chi_s \,ds  \right| \xi \right\rangle + C e^{-R/2} \|\xi\|^2 \,.
\end{equation}
In order to bound $S_2$, it thus suffices to consider
functions  $\xi$ whose Fourier transform $\hat \xi(q,\hat p_{12})$ is  supported on a set
\begin{equation}\label{supp}
e^{s-L} \leq  K(q,\hat p_{12})   \leq e^{s+L} 
\end{equation}
for some $s\in \R$.

\subsection{Step 3. Simple upper bound on $S_2$}
Using the change of variables 
\begin{equation}
	\label{eq:changeofvars}
	P=p_1+p_2 \,,\quad Q=p_3+p_4\,,\quad p = (p_1-p_2)/2\,,\quad q=(p_3-p_4)/2\,,
\end{equation}
we write
\begin{align} \nonumber
	 \langle \xi \left| S_2 \right| \xi\rangle 
	&=
	\int \frac{  \widehat \xi(P, Q/2+q,Q/2-q, p_5,\dots, p_N)^*   \widehat \xi( Q, P/2+p, P/2 -p ,p_5,\dots,p_N )}{ \frac 12 (P^2+Q^2) + 2 (p^2+q^2) + \sum_{j=5}^N p_j^2 + 1 } \\ & \qquad \times dP dp dQ dq dp_5 \cdots dp_N \,.
	\label{37a}
\end{align}
For fixed $P,Q,p_5,\ldots,p_N$, a simple Cauchy--Schwarz inequality (using the positivity of the operator on $L^2(\R^2)$ with integral kernel $1/(|x|^2+|y|^2+\lambda^2)$; see Lemma \ref{lem:1}) implies that
\begin{align}\nonumber
	& \int \frac{  \widehat \xi(P, Q/2+q,Q/2-q, p_5,\dots, p_N)^*   \widehat \xi( Q, P/2+p, P/2 -p ,p_5,\dots,p_N )}{ \frac 12 (P^2+Q^2) + 2 (p^2+q^2) + \sum_{j=5}^N p_j^2 + 1 }\, dp dq \\ \nonumber
	& \leq \sqrt{ \int \frac{  \widehat \xi(P, Q/2+q,Q/2-q, p_5,\dots, p_N)^*   \widehat \xi( P, Q/2+p, Q/2 -p ,p_5,\dots,p_N )}{ \frac 12 (P^2+Q^2) + 2 (p^2+q^2) + \sum_{j=5}^N p_j^2 + 1 }\, dp dq} \\
	& \quad \times \sqrt{ \int \frac{  \widehat \xi(Q, P/2+q,P/2-q, p_5,\dots, p_N)^*   \widehat \xi( Q, P/2+p, P/2 -p ,p_5,\dots,p_N )}{ \frac 12 (P^2+Q^2) + 2 (p^2+q^2) + \sum_{j=5}^N p_j^2 + 1 }\, dp dq} \,.
\end{align}
We integrate this inequality with respect to $P$ and $Q$ and apply another Cauchy--Schwarz inequality with the two square roots on the right side. After integrating with respect to $p_5,\ldots,p_N$, we arrive at the bound 
\begin{align}\nonumber 
 \langle \xi \left| S_2 \right| \xi\rangle 
 & \leq  \int \frac{ \widehat \xi(P, Q/2+q,Q/2-q, p_5,\dots, p_N)^*   \widehat \xi( P, Q/2+p, Q/2 -p ,p_5,\dots,p_N )}{ \frac 12 (P^2+Q^2) + 2 (p^2+q^2) + \sum_{j=5}^N p_j^2 + 1 } \\ & \qquad \times dP dp dQ dq dp_5 \cdots dp_N \,.
\label{37}
\end{align}

Moreover, for $\hat \xi \geq 0$ we can further bound the denominator from below by $ \frac 12 (P^2+Q^2) + p^2+q^2 + \sum_{j=5}^N p_j^2 + 1$, which can be written as 
\begin{align}\nonumber
&\frac 1{\frac 12 (P^2+Q^2) + p^2+q^2 + \sum_{j=5}^N p_j^2 + 1} \\ & = 
\int_0^\infty e^{-\frac t2 \left( K(P,Q/2+q,Q/2-q,p_5,\dots,p_N) + K(P,Q/2+p,Q/2-p,p_5,\dots,p_N)\right)} \,dt \,.
\end{align}
The resulting  expression is thus given by 
\begin{equation}
(2\pi)^2 \int_0^\infty \left\langle \xi \left| e^{-t K/2} \delta_{34} e^{-t K/2} \right| \xi \right\rangle dt \,,
\end{equation}
where $\delta_{34}$ denotes multiplication by $\delta(x_3-x_4)$. That is, we have shown that
\begin{equation}\label{40}
\langle \xi \left| S_2 \right| \xi\rangle  \leq (2\pi)^2 \int_0^\infty \left\langle \xi \left| e^{-t K/2} \delta_{34} e^{-t K/2} \right| \xi \right\rangle dt
\end{equation}
if $\hat \xi \geq 0$. 

\subsection{Step 4. Mollifying the interaction} 
We shall now replace the interaction $\delta(x_3-x_4)$ in \eqref{40} by a mollified interaction $w(x_3-x_4)$. Assume that $w$ is a smooth, nonnegative, compactly supported, even function on $\R^2$, with  $\int w = 1$ and $\int_{\R^2} w(x) e^{ipx} dx  \geq 1 - C_w |p|^2$ for all $p\in \R^2$. For $\hat \xi$ supported on $\vec p$ satisfying \eqref{supp},  all momenta $p_3,\dots,p_N$ in the support of $\hat \xi$ satisfy $|p_j| \leq e^{(s+L)/2}$. In particular, $|p-q| \leq 2 e^{(s+L)/2}$ in the integrand on the right hand side of \eqref{37}. Consequently, with $w^\kappa(x) = \kappa^{2} w(\kappa x)$ for some $\kappa>0$
\begin{equation}
\int_{\R^2} w^\kappa (x) e^{i(p-q)x} dx  \geq 1 - C_w \kappa^{-2}|p-q|^2 \geq 1-4 C_w \kappa^{-2} e^{s+L} \,.
\end{equation}
We thus conclude that 
\begin{equation}\label{42}
\left\langle \xi \left| e^{-t K/2} \delta_{34} e^{-t K/2} \right| \xi \right\rangle \leq \left( 1 - 4 C_w \kappa^{-2} e^{s+L} \right)^{-1} \left\langle \xi \left| e^{-t K/2} w^\kappa_{34} e^{-t K/2} \right| \xi \right\rangle
\end{equation}
for any $\xi$ with $\hat \xi \geq 0$ supported in \eqref{supp}. Here $w^\kappa_{34}$ is short for multiplication by $w^\kappa(x_3-x_4)$. 

\subsection{Step 5. Ground state energy in the mean-field limit.} 
With $w$ as above, pick $0<a<a_*$ (defined in \eqref{LGN}), and let 
\begin{equation}\label{def:hna}
E_n(a) = \infspec \left[ \sum_{i=1}^n \left( -\Delta_i\right) - \frac a {n-1} \sum_{1\leq i<j\leq n} w(x_i -x _j)  \right]
\end{equation}
for $n\geq 2$, where the operator acts in $L^2_{\rm sym}(\R^{2n})$. Then, by a simple rescaling, we have the bound 
\begin{equation}
\sum_{1\leq i<j\leq n} w^\kappa (x_i -x _j)  \leq \frac {n-1} a \left( - \kappa^2 E_n(a) +   \sum_{i=1}^n \left( -\Delta_i\right) \right)
\end{equation}
for any $\kappa>0$. We shall apply this in \eqref{42}, for $n = N-2$. By the bosonic symmetry of $\xi$ in the last $N-2$ variables, 
\begin{align}\nonumber
&  \int_0^\infty \left\langle \xi \left| e^{-t K/2} w^\kappa_{34} e^{-t K/2} \right| \xi \right\rangle dt 
\\ &\leq   \frac {2} {a(N-2)}  \left( - \kappa^2 E_{N-2}(a) \left\langle \xi \left| \frac{1}{K} \right|\xi\right\rangle +  \sum_{j=3}^N  \left\langle \xi \left| \frac{-\Delta_j}{K} \right|\xi\right\rangle  \right)
\nonumber \\
&\leq   \frac {2} {a(N-2)}  \left( 1 - e^{L-s} \kappa^2 E_{N-2}(a)  \right),\label{45}
\end{align}
where we used the lower bound on $K$ in \eqref{supp} in the first term on the right hand side.

\subsection{Step 6. Combining the bounds}  
A combination of \eqref{40}, \eqref{42} and \eqref{45} shows that, for any normalized $\xi$ with $\hat \xi \geq 0$ supported in \eqref{supp} for some $s \in \R$, 
\begin{equation}
\langle \xi | S_2 | \xi \rangle  \leq   \frac {2 (2\pi)^2} {a(N-2)}   \left( 1 - 4 C_w \kappa^{-2} e^{s+L} \right)^{-1}  \left( 1 - e^{L-s} \kappa^2 E_{N-2}(a)  \right)  
\end{equation}
for any $\kappa>0$. Given $\eps>0$, we shall choose $\kappa$ such that $\kappa^{-2} e^{s+L} = \eps$, hence this becomes 
\begin{equation}
\langle \xi | S_2 | \xi \rangle  \leq   \frac {2 (2\pi)^2} {a(N-2)}   \left( 1 - 4 C_w \eps \right)^{-1}  \left( 1 - e^{2 L} \eps^{-1}  E_{N-2}(a)  \right),  
\end{equation}
which is now independent of $s$. 
Further combining this with \eqref{35}, we finally conclude that
\begin{equation}
\| S_2\| \leq \frac {2 (2\pi)^2} {a(N-2)} \left( 1 - C(R/L)^2\right)^{-1}  \left( 1 - 4 C_w \eps \right)^{-1}  \left( 1 - e^{2 L} \eps^{-1}  E_{N-2}(a)  \right)   + C e^{-R/2} \,.
\end{equation}
This bound holds for all $R,L>0$ such that $R/L$ is sufficiently small and for all $\epsilon>0$. For the first error term to be small, we choose $R/L = \eps^{1/2}$. For the last one we need $e^{-R/2} = \eps/N$. This leaves us with showing that 
\begin{equation}
 \eps^{-1} (N/\eps)^{4 \eps^{-1/2} }  E_{N-2}(a) 
\end{equation}
is small for large $N$, i.e., that $E_{N-2}(a)$ goes to zero faster than any power of $N$. This will be deduced in the final step from the work in \cite{JV}. Accepting this for the moment, we deduce that
\begin{equation}
\|S_2\| \leq \frac{2(2\pi)^2}{a(N-2)} \left( 1+ o(1) \right)
\end{equation}
as $N\to\infty$.

Combining this with the operator inequality $S_0\geq 0$ and the norm bound on $S_1$ from Step 1, we deduce that
\begin{equation}
N^{-1} t_N \geq -2(1-2/N) \|S_1\| - \frac12 (N-2)(1-3/N)\|S_2\|
\geq - (2\pi)^2 a^{-1} (1+o(1)) \,.
\end{equation}
Since $a<a_*$ is arbitrary, this implies
\begin{equation}\label{eq:lowerboundfinal}
\liminf_{N\to\infty} \frac{t_N}{N} \geq - \frac{(2\pi)^2}{a_*} \,,
\end{equation}
which is the claimed lower bound.

\subsection{Step 7. Rough bound on ground state energy} In \cite[Thm.~2]{JV}, it was shown that for any $0<a<a_*$ and  $\kappa_n = e^{n^\beta}$ for some $\beta < 1$, 
\begin{equation}\label{eq:jv}
 \sum_{i=1}^n \left( -\Delta_i + V(x_i) \right) - \frac a {n-1} \sum_{1\leq i<j\leq n} w^{\kappa_n}(x_i -x _j)  \geq - n K 
\end{equation} 
for some constant $K$ depending only on $a$, $w$, $V$ and $\beta$. 
Here $V$ is a confining potential growing like a power of $|x|$. 
After rescaling by $\kappa_n$, this becomes
\begin{equation}\label{51}
\sum_{i=1}^n \left( -\Delta_i + \kappa_n^{-2} V(x_i/\kappa_n) \right) - \frac a {n-1} \sum_{1\leq i<j\leq n} w(x_i -x _j)    \geq - n \kappa_n^{-2} K \,.
\end{equation}

To get a lower bound on $E_n(a)$, we can use IMS localization. Let $H_n(a)$ denote the operator on the right-hand side of \eqref{def:hna}. It is convenient to view it as an operator on all of $L^2(\R^{2n})$, it is well-known that the infimum of its spectrum is unchanged upon restriction to the bosonic subspace. IMS localization implies that for any $\ell >0$ there exist $z_i\in \R^2$ and  a $\psi \in L^2(\R^{2n})$ supported in $\bigtimes_{i=1}^n B_\ell(z_i)$ such that $\langle \psi | H_n(a) |\psi\rangle \leq E_n(a) + n C \ell^{-2}$, with $B_\ell(z)\subset \R^2$ denoting the ball of radius $\ell$ centered at $z\in\R^2$. Since $w$ is compactly supported by assumption, an interaction term $w(x_i-x_j)$ will  be zero if $|z_i-z_j| > 2 \ell + d$, with $d$ the diameter of the support of $w$. We thus slightly enlarge the balls $B_\ell$ to $B_{\ell + d/2}$, and write their union as a disjoint union of $m$ clusters, with $1\leq m \leq n$, each consisting of finitely many balls: $\cup_{i=1}^n B_{\ell+d/2}(z_i) = \cup_{j=1}^m S_j$, the latter sets being mutually disjoint. Since there are no interactions between particles in different clusters, we have 
\begin{equation}
\left\langle\psi | H_n(a) | \psi \right\rangle = \sum_{j=1}^m \langle \psi | H^{S_j} | \psi\rangle
\end{equation}
where $H^{S_j}$ is the Hamiltonian for the particles in cluster $S_j$. 

Each cluster \(S_j\) is a connected union of $1\leq m_j \leq n$ balls of radius \(\ell+d/2\), and is therefore contained in a ball \(B_{D}(w_j)\) for some $w_j\in\R^2$, where \(D:=n(\ell+d/2)\). We  now translate all the coordinates in $S_j$ by $w_j$, i.e., center all the balls at the origin. This leaves the interactions within clusters invariant, but may introduce additional attractive interactions between  particles in different clusters. The translated wave function $\tilde \psi$  hence satisfies $\langle\tilde \psi | H_n(a) |\tilde\psi \rangle \leq \langle \psi | H_n(a) | \psi\rangle$. Since by construction $\tilde\psi$ is supported in $\bigtimes_{i=1}^n B_D(0)$, we can add  one-body potential terms $D^{-2}\sum_{k=1}^n V(x_k/D)$ to the Hamiltonian $H_n(a)$ at the energy cost of at most $n D^{-2} \sup_{B_{1}(0)} V$.  Using \eqref{51} with $\kappa_n = D$, i.e., choosing $\ell = \kappa_n/n -d/2$, we conclude that 
\begin{equation}
E_n(a) \geq  - \frac{Cn}{\ell^2} - \frac n{D^2} \left( K + \sup_{B_{1}(0)} V\right) \geq - C n^3 \kappa_n^{-2}
\end{equation} 
in the case $\kappa_n \geq d n$. 
This 
completes the proof. 


\subsection*{Acknowledgments}
R.S. is grateful for inspiring discussions with Artem Volosniev. R.L.F. acknowledges partial support by the German Research Foundation grants EXC-2111-390814868 and TRR 352–Project-ID 470903074, as well as through the ChatGPT for Academic Researchers program of OpenAI. R.S. acknowledges partial support by the Austrian Science Fund (FWF) grant DOI: 10.55776/I6427 (as part of the SFB/TRR 352). 

\subsection*{AI declaration}
The entire text of the manuscript was written by the authors without assistance of AI tools. Crucial ideas for the proof of the lower bound in Section \ref{lb} were obtained with the help of GPT-6 Astra. Claude Opus 5.5 was used for proof-reading.




\begin{thebibliography}{14}
\bibitem{BCEZ} F. Bentosela, R.M. Cavalcanti, P. Exner, V.A. Zagrebnov, \textit{Anomalous electron trapping by localized magnetic fields}, J. Phys. A \textbf{32},  3029--3039 (1999).
\bibitem{logdF1} F.G. Brand\~{a}o,  A.W. Harrow, {\it Quantum de Finetti theorems under local measurements
with applications}, Commun. Math. Phys. {\bf  353}, 469--506 (2017).
\bibitem{DA} G. Dell-Antonio, R. Figari, A. Teta, {\it Hamiltonians for systems of $N$ particles interacting through point
interactions},  Ann. Inst. Henri Poincar\'e {\bf 60}, 253--290 (1994).
\bibitem{DR} J. Dimock, S.G. Rajeev, {\it Multi-particle Schr\"odinger operators with point interactions in the plane}, J. Phys. A: Math. Gen. {\bf 37}, 9157 (2004).
\bibitem{GH} M. Griesemer, M. Hofacker, {\it From Short-Range to Contact Interactions in Two-dimensional Many-Body Quantum Systems}, Ann. Henri Poincar\'e {\bf 23}, 2769--2818 (2022).
\bibitem{HS} H.-W. Hammer, D.T. Son, {\it Universal Properties of Two-Dimensional Boson Droplets}, Phys. Rev. Lett. {\bf 92}, 250408 (2004).
\bibitem{JV}  L. Junge, F.L.A. Visconti, {\it Derivation of Hartree theory for two-dimensional attractive Bose gases in almost Gross--Pitaevskii regime},
J. Math. Pures  Appl. {\bf  205}, 103800 (2026).
\bibitem{LNR} M. Lewin, P.T. Nam, N. Rougerie, \textit{Derivation of Hartree’s theory for generic mean-field Bose systems}, Adv. Math. \textbf{254},  570--621 (2014).
\bibitem{logdF2} K. Li, G. Smith, {\it Quantum de Finetti theorem under fully-one-way adaptive measurements}, Phys. Rev. Lett. {\bf 114}, 160503 (2015).
\bibitem{R} S.G. Rajeev, {\it A Condensation of Interacting Bosons in Two Dimensional Space}, unpublished, preprint arXiv:9905120
\bibitem{R1} N. Rougerie, {\it Non linear Schr\"odinger limit of bosonic ground states, again}, 
Confluentes Math. {\bf 12}, 69--91  (2020). 
\bibitem{R2} P.T. Nam, N. Rougerie, {\it Improved stability for 2D attractive Bose gases}, J. Math. Phys. {\bf 61}, 021901 (2020). 
\bibitem{S} B. Simon, {\it The Bound State of Weakly Coupled Schr\"odinger Operators in One and Two Dimensions}, Ann. Phys. {\bf 97}, 279--288 (1976).
\bibitem{tms} G.V. Skorniakov, K.A. Ter-Martirosian, {\it Three body problem for short range forces. I. Scattering of low
energy neutrons by deuterons},  Sov. Phys. JETP {\bf 4}, 648--661 (1957).
\bibitem{AV}  M. Suchorowski, F. Brauneis, H.-W.  Hammer, M. Tomza, A.G. Volosniev, 
{\it Generalized Gross--Pitaevskii Equation for 2D Bosons with Attractive Interactions},
Phys. Rev. Lett. {\bf 136}, 243402 (2026).
\bibitem{W} M.I. Weinstein, \textit{Nonlinear Schr\"odinger equations and sharp interpolation estimates}, Comm. Math. Phys. \textbf{87}, 567--576 (1983).
\end{thebibliography}
\end{document}